# Dielectric Behavior as a Screen in Rational Searches for Electronic Materials: Metal Pnictide Sulfosalts

Xin He,[1] David J. Singh,[2,*] Patsorn Boon-on,[3] Ming-Way Lee[3,*] and Lijun Zhang[1,*]

[1]State Key Laboratory of Superhard Materials, Key Laboratory of Automobile Materials of MOE, and School of Materials Science, Jilin University, Changchun 130012, China

[2]Department of Physics and Astronomy, University of Missouri, Columbia, Missouri 65211-7010, United States

[3]Institute of Nanoscience and Department of Physics, National Chung Hsing University, Taichung 402, Taiwan

*Authors to whom correspondence should be addressed: singhdj@missouri.edu, mwl@phys.nchu.edu.tw, lijun_zhang@jlu.edu.cn

## Abstract

Dielectric screening plays an important role in reducing the strength of carrier scattering and trapping by point defects for many semiconductors such as the halide perovskite solar materials. However, it was rarely considered as a screen to find new electronic semiconductors. We performed a material search study using the dielectric properties as a screen to identify potential electronic materials in the class of metal-pnictide ternary sulfosalts, containing Bi or Sb. These salts are basically ionic due to the electronegativity difference between the S and both the metal and pnictogen elements. However, we do find significant cross-gap hybridization between the S p-derived valence bands and pnictogen p-derived conduction bands in many of the materials. This leads to enhanced Born effective charges, and in several cases, highly enhanced dielectric constants. We find a chemical rule for high dielectric constants in terms of the bond connectivity of the pnictogen-chalcogen part of the crystal structure. We additionally find a series of compounds with low effective mass, high dielectric constant and other properties that suggest good performance as electronic materials and also several potential thermoelectric compounds. Experimental optical data and solar conversion efficiency are reported for Sn-Sb-S samples, and results in accord with predicted good performance are found. The results illustrate the utility of dielectric properties as a screen for identifying complex semiconductors.

## Introduction

Important discoveries in the last several years have led to renewed interest in complex semiconductors, for example, the halide perovskite solar materials.[1–5] This previously overlooked family of materials exhibited unexpected high photovoltaic performance, as their unique intrinsic properties give rise to high visible-light capture capability,[6,7] facile photoinduced carrier separation,[8,9] and efficient ambipolar carrier transport.[10–12] These discoveries also represent a new approach in semiconductor physics, in particular, a focus on materials whose high performance is due to intrinsic factors, rather than exquisite control of crystalline perfection and the elimination of defects, and has led to a resurgence in the search for new non-traditional semiconductor materials.[13] This also provides an important direction for semiconductor applications where defects cannot be eliminated even in principle, a case in point being thermoelectrics, where heavy doping and high mobility are simultaneously required, along with other properties such as high thermopower that preclude a simple resolution, e.g. in terms of strong band edge dispersion and low effective mass, that have been useful guides in other semiconductor applications.[14] Instead ways of avoiding carrier scattering from point defects are required.[15,16] The mechanisms underlying the performance of materials such as the halide semiconductors are not yet fully elucidated, but it has been argued that dielectric screening plays an important role in preventing carrier scattering and trapping by point defects.[17–20] The static dielectric constants of this family of materials are as high as 70 and above.[21-22] This gives rise to ultralong carrier diffusion length and efficient carrier transport,[10-12] which is thought to greatly facilitate optoelectronic conversion efficiency.

Searching for complex semiconductors having simultaneously high dielectric screening and favorable electronic structure may be a promising direction to discover new functional materials for electronic and optoelectronic applications, in particular complex semiconductors with high defect tolerance. This is not to say that all important semiconductors can be understand and found in this way. For example, $Cu_2ZnSn(S,Se)_4$ is an important multinary solar material that has demonstrated good properties and efficiency of ~13%,[23] with a modest static dielectric constant of ~15.[24] Nonetheless, the new paradigm exemplified by the halides has led to exciting new materials and high performance devices.

Here, we perform such a rational search study based on first principles materials screening to illustrate the approach. Our focus is a family of materials with large combinatorial space of candidates, the metal (M) – pnictide (Sb, Bi) - S ternary sulfosalts, for which we systematically investigate compounds reported in the Inorganic Crystal Structure Database (ICSD).[25] In particular, we performed detailed first-principles calculations for 116 reported ternary compounds. This group of compounds is already known to include some good semiconductors, including, for example, $NaSbS_2$, which is a good solar absorber and potential thermoelectric,[26–28] and the earth-abundant, possibly biocompatible semiconductor, $NaBiS_2$.[29] The dielectric screening behaviors and intrinsic properties (e.g., electronic structure, effective masses of electron/hole, optical absorption and thermoelectric properties) of M-Sb/Bi-S are calculated, and the trends are analyzed. We find several compounds with properties consistent with excellent semiconducting behaviors, including high dielectric constant, favorable transport effective mass values, and suitable band gaps for particular applications. We also identify a useful chemical rule for high dielectric constant in terms of structural motif, in particular the bond connectivity of the pnictogen-chalcogen part of the crystal structure. Our preliminary experimental results on Sn-Sb-S compounds for solar absorbers supports the theoretical predictions.

The basis for our choice of using dielectric constant as a screen and for focusing on pnictogen sulfosalts is illustrated in Figure 1. Figure 1a shows a plot of the product of mobility and effective mass for known good semiconductors vs static dielectric constant. The product of mobility and effective mass is average inversely proportional to the carrier scattering rate; that is, it removes the well-known inverse relation between effective mass and mobility. It represents a crucial factor in carrier transport in semiconductors. As seen there is a substantial correlation even in these classic semiconductors. Furthermore, it has been demonstrated by experiment that there is a clear connection between dielectric constant and mobility in diverse materials, including oxide electronic materials[30] and layered transition metal dichalcogenides.[31] Figure 1b shows our calculated electronic structure of $PbBi_2S_4$, which is one of the M-Sb/Bi-S compounds studied (see below for details). This is a material that was previously identified as a good potential solar absorber based on band structure and optical considerations.[30] The electronic structure is characteristic of a salt, $Pb^{2+}(Bi^{3+})_2(S^{2-})_4$, as may be expected from the electronegativity differences, and additionally shows strong cross-gap hybridization between the S p states comprising the valence band and pnictogen p states comprising the conduction band. As illustrated in Figure 1b, this provides a mechanism for strongly enhanced Born effective charges, similar to perovskite ferroelectrics.[33] The enhanced Born effective charges in the ternary M-Sb/Bi-S sulfosalts are expected to lead to a dominant lattice contribution in the static dielectric constant.

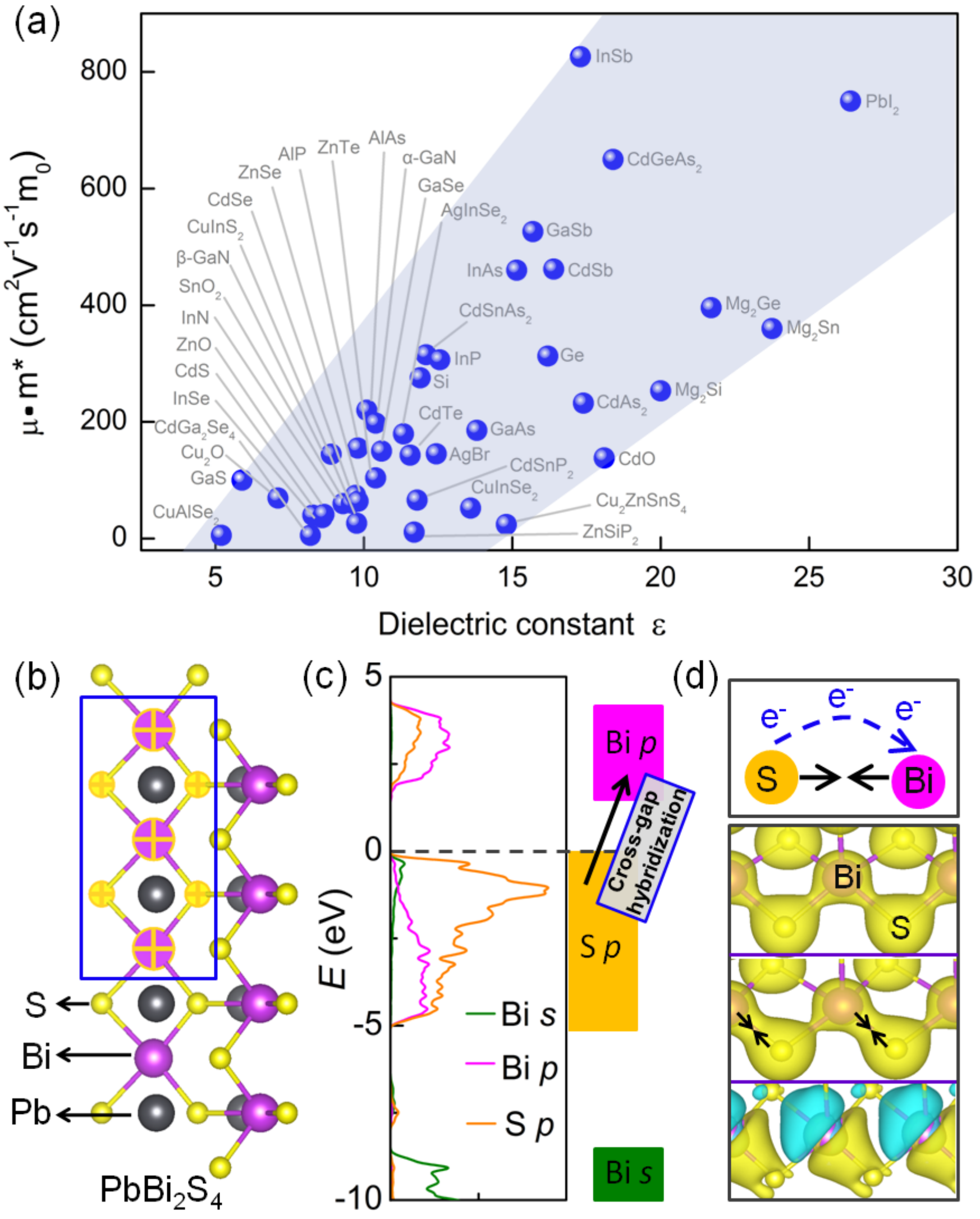

**Figure 1.** (a) Product of mobility and effective mass vs dielectric constant from experiment for known semiconductors. The compounds shown are $CuAlSe_2$,[34] GaS, , $Cu_2O$, ZnO, $CdGa_2Se_4$,[35] CdS, InSe, ZnSe,[36] β-GaN, InN,[37] $SnO_2$, AlP, AlAs, $CuInS_2$,[38] ZnTe, α-GaN, CdTe,[36,39] GaSe, $AgInSe_2$,[40] $ZnSiP_2$, $CdSnP_2$, diamond structure Si, $CdSnAs_2$, CdSe, AgBr, InP, GaAs,[36] $CdSnP_2$, $CuInSe_2$,[41] $Cu_2ZnSnS_4$,[24,42] GaSb, Ge, CdSb, InSb, $CdAs_2$, CdO,[43] $CdGeAs_2$, $Mg_2Si$, $Mg_2Ge$, $Mg_2Sn$, $PbI_2$. Data is from O. Madelung,[44] except where indicated. (b) Crystal structure of a studied compound, $PbBi_2S_4$. (c) Projected electronic density of states of $PbBi_2S_4$ showing nominal Bi p conduction bands and S p valence bands with cross-gap hybridization between them. As illustrated schematically in (d, upper panel), this leads to a transfer of charge from S to Bi due to increased hybridization when the bond is compressed. This corresponds to an additional effective charge motion, i.e., enhanced Born effective charges (Z*) and resulting high dielectric constants. The lower panels of (d) shows the charge density before distortion, after and the difference (green and blue are positive and negative).

## Computational Methods

The present calculations are performed within density-functional theory (DFT).[45] Two electronic structure methods were employed: the projector augmented wave (PAW) method as implemented in the VASP code,[45] because of the efficient relaxations and calculations of dielectric properties that are possible in that method, and the general potential linearized augmented planewave (LAPW) method as implemented in the WIEN2K code,[47] due to the highly accurate, all electron electronic structures, suitable for transport and optical calculations that can be obtained with this method. We used the standard Perdew-Burke-Ernzerhof (PBE) generalized gradient approximation (GGA)[48] to relax the atomic coordinates by total energy minimization, while fixing the lattice parameters to experimental values. This is important because errors on the lattice parameters can have strong effects on lattice dynamics and dielectric behavior. Born effective charge and dielectric tensor calculations were performed using density functional perturbation theory through the VASP code. In all of above calculations the convergence limit was set to $1\times 10^{-4}$ eV Å$^{-1}$ for the force and the $1\times 10^{-5}$ eV atom$^{-1}$ for the energy.

The electronic structures were then obtained with the LAPW method,[49] as implemented in the WIEN2K code. We used LAPW sphere radii of 2.2 bohr for all elements. Standard density functionals, such as the PBE GGA, generally underestimate semiconductor band gaps. In order to correct this, we used the modified Becke-Johnson (mBJ) potential due to Tran and Blaha for these calculations.[50] This potential generally improves the accuracy of the gap values relative to experiment for semiconductors, and is computational efficient even for dense sets of k-points, as needed in transport calculations. [51,52] Spin orbit coupling (SOC) was included, since it is potentially important for electronic structures of heavy p-electron systems, as in the Bi compounds that we study (note however that spin-orbit has a much smaller effect on total energies and other properties that depend on averages over electronic states, such as crystal structures, phonons and Born charges).[18] All thermoelectric properties reported were calculated using BoltzTraP code.[53] Transport calculations require a dense sampling of the Brillouin zone. We used at least 5000 k-points in the Brillouin zone from the first principles calculations. These formed the basis of the interpolation implemented in BoltzTraP. Thermoelectric electronic fitness functions (EFF) were obtained using the transM code.[53] To evaluate material-intrinsic solar cell efficiency, the "spectroscopic limited maximum efficiency (SLME)" based on the improved Shockley-Queisser model is calculated.[55] This measure, although it has some limitations, e.g. due to device assumptions, is a useful screen that is applicable in high throughput searches, and emphasizes the importance of high absorption for solar cell materials.[55] While here we used high dielectric constant as a screen and suitable band gap, transport effective mass, and SLME as secondary screens to illustrate the approach, other secondary screens could also be used in a high throughput setting.

## Results and Discussion

We investigate 116 ternary sulfosalts in M-Sb/Bi-S system, and selected for further study those compounds that were found to have semiconducting band gaps at the PBE level of theory. These are the 78 identified semiconductors. Some of the other compounds may also have small band gaps at higher levels of theory, but they are not further studied here. The calculated band gaps as obtained with the mBJ potential, including SOC, of are listed in Table S1, compared with available experimental results, along with the list of compounds that do not have a gap at the PBE level. As seen, the calculated gap values are in reasonable accord with corresponding experimental data. Errors are generally in the range of 0.1 - 0.2 eV, and larger discrepancies are present for Cu compounds, e.g. $Cu_3BiS_3$. While these errors may be significant for some applications, they are in the range of experimental error for many of these complex semiconductors, and represent good agreement for current first-principles calculations based on nonfitted methods applicable to high throughput searches. The calculated band gaps, pnictogen Born effective charge, and effective mass of electrons and holes are plotted vs the calculated static dielectric constants in Figure 2. For this purpose we show the direction averages, specifically the average dielectric constant and the inverse of the average value of the inverse effective mass tensor, determined from transport calculations, i.e. the average transport effective mass (see Li et al.[57] for details).

We begin with the discussion of the identified chemical trends. As shown in the Figure 2a, the calculated band gaps of these M-Sb/Bi-S compounds cover a wide range from ~0.3 to ~3.5 eV. The dielectric constants, as obtained from PBE calculations, are large for many of the compounds. This reflects the generic expectation of cross-gap hybridization in such compounds (see Figure 1). However, there are large variations in dielectric constant between these materials, trends for which are discussed below. High dielectric constants are expected to lead to screening of defects and suppression of point defect scattering and carrier trapping. There is a general trend of decreasing band gap with increasing dielectric constant, as seen in Figure 2a. However, importantly, this is not universal. In particular, it is possible to find materials in any band gap range that is obtainable in this chemistry, while simultaneously having high dielectric constant. Furthermore, there is a clear connection between the Born effective charge of the pnictogen (Sb or Bi) and dielectric constant, as shown in Figure 2b. This is a consequence of the origin of the high dielectric constants in the lattice part, associated with the cross-gap hybridization as mentioned. Finally, in most cases the electron effective masses are lower than those of holes, but there is no obvious relationship between the effective masses and the dielectric constant (Figure 2c). This means that again it is possible to find high dielectric constant materials, with low effective mass, especially for electrons. Thus based on trends, the use of dielectric constant as a search criterion has considerable utility. Moreover, strikingly, many of these compounds possess electron or hole effective masses lower than or around the rest mass of electrons ($m_0$), indicating the possibility of good carrier mobility. The large span of the band gaps along with low electron or hole effective mass, and high dielectric constant, suggests opportunities for various optoelectronic applications based on compounds in this M-Sb/Bi-S class of compounds.

The compounds that have a desirable combination of small electron/hole effective mass ($m_e$ or $m_h < m_0$), band gap Eg >1 eV and large dielectric constant ($\varepsilon_0 > 20$) are shown in Figure 2d. As seen, there are a substantial number of such compounds. We find that these compounds have a structural commonality. In particular, we find that the connectivity of the pnictogen–S sublattice is key. Those compounds that have connectivity in at least one dimension usually have high static dielectric constant, as illustrated by the structures shown in Figure S2. Those compounds that have disconnected pnictogen–S anionic units on the other hand tend to have lower dielectric constant, presumably because they do not have the possibility of a polar symmetry lowering associated with off-centering corresponding to displacement of a pnictogen between two or more S atoms, analogous to ferroelectric materials, and as represented in the schematic on the right side of Figure 1b. The essential ingredients, besides crystal structures, which have pnictogens placed symmetrically or nearly symmetrically between S atoms (the connectivity of the pnictogen–S sublattice), are the ionic nature of these salts and the cross-gap hybridization between S p and pnictogen p orbitals.

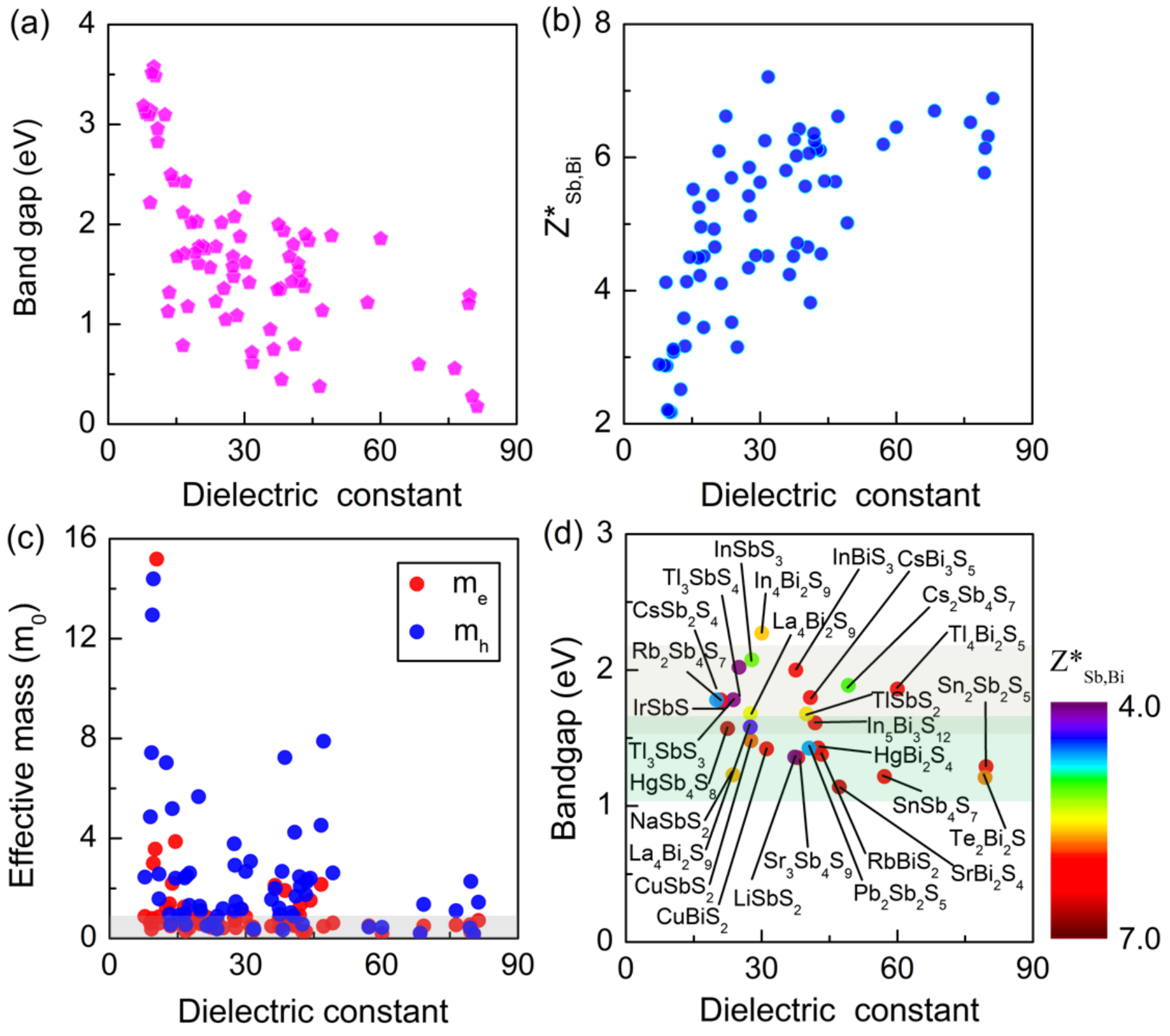

**Figure 2.** (a) Calculated Band gaps, (b) Pnictogen Born effective charges and (c) electron ($m_h$) and hole ($m_e$) effective mass vs dielectric constant of M-Sb/Bi-S semiconductors. (d) Band gaps vs dielectric constant of compounds have small electron/hole effective mass ($m_e$ or $m_h$ < $m_0$) and large dielectric constant ($\varepsilon_0 > 20$), the color of the circles indicate the pnictogen Born effective charges.

We now turn to compounds with favorable properties identified in our search. We focus on potential optoelectronic or thermoelectric materials. A suitable optical band gap is a primary requirement for specific optoelectronic applications, for example, in the case of solar absorbers, band gaps near the maximum in the Shockley-Queisser curve are desired.[58] Thus, compounds with optical band gaps in range of ~1.0-1.7 eV may be suitable for application as solar absorbers (note the Shockley-Queisser curve is relatively flat in the peak region). Those with band gaps in the range of 1.5-2.3 eV are potential semiconductors for room-temperature radiation detection. We identify 28 compounds with an electron or hole effective mass lower than the free electron mass ($m_0$), static dielectric constant $\varepsilon_0$>20, and band gap 2.3 eV > Eg > 1.0 eV among the compounds studied. These calculated indirect and direct band gaps and SLME for a film thickness of 1$\mu$m based on the improved Shockley-Queisser model are shown in Figure 3a. The calculated optical absorption spectra and SLME are shown in Figure 3 and Figure S1. All these compounds have high absorption coefficient (more than $10^5$ $cm^{-1}$). $Sn_2Sb_2S_5$,[59] $SnSb_4S_7$,[60] $CuBiS_2$, $CuSbS_2$,[61] and $NaSbS_2$,[27] were previously studied experimentally and in several cases ($CuBiS_2$, $CuSbS_2$ and $NaSbS_2$) found to have reasonable optoelectronic properties in experiment studies. They have been shown to have large absorption and SLME in our calculation. The identification of these compounds by our screen provides a measure of validation. We also find some compounds that have not been studied experimentally but may be high-performance optoelectronic materials based on the results of our screening. For instant, the SLME of $Sr_3Sb_4S_9$ and $SrBi_2S_4$ were up to 30% under 1 μm thickness. We chose 13 compounds which have SLMEs larger than 25% and show their optical absorption spectrum and SLME in Figure 3b,c. Those band structure and PDOS are shown in Figure S2-S5.

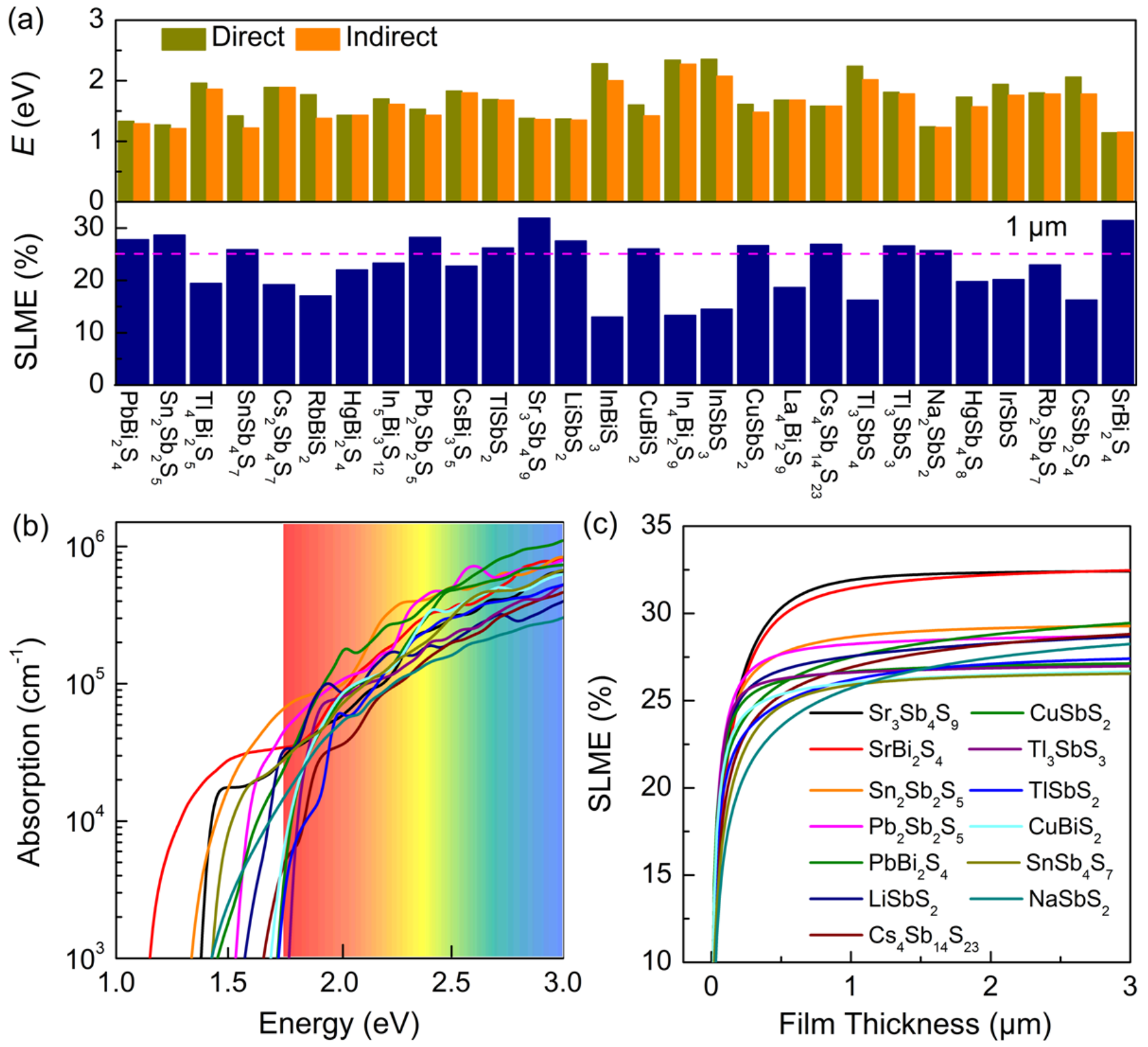

**Figure 3.** (a) Calculated indirect and direct band gaps, and "spectroscopic limited maximum efficiency (SLME)" under film thickness 1$\mu$m based on the improved Shockley-Queisser

model of 28 compounds that have band gap 2.3 eV > Eg >1 eV, small electron/hole effective mass ($m_e$ or $m_h$ < $m_0$) and large dielectric constant ($\varepsilon_0$ > 20). (b) Calculated absorption spectra and SLME of the 13 compounds that have SLME values larger than 25%. (c) Thickness dependent SLME.

As seen, two of the Sn-Sb-S phases, specifically $Sn_2Sb_2S_5$ and $SnSb_4S_7$ show favorable properties. We synthesized ternary nanoparticles of Sn-Sb-S using a two-step solution processing process: sequential ionic layer adsorption reaction (SILAR), followed by annealing in flowing $N_2$ gas.

The synthesis procedure is as follows. Binary metal sulfide Sn-S nanoparticles were grown on a mesoporous $TiO_2$ electrode. Then binary metal sulfide Sb-S nanoparticles were grown on top of the Sn-S nanoparticles. Postannealing was done to convert the double-layered Sn-S/Sb-S structure into the ternary Sn-Sb-S semiconductor. The Sn-S SILAR cycles were done by dipping the $TiO_2$ electrodes for 30 s into a 0.1 M, 300 K $SnCl_2$ ethanol solution, followed by an ethanol rinse, then dipping in a 0.1 M 300 K $Na_2S$ methanol solution also for 30 s, and finally a methanol rinse and air-dry. The Sb-S SILAR cycles were done also at 300 K, similar to the Sn-S cycles, but using 0.1 M $SbCl_3$ ethanol and 0.1 M $Na_2S$ methanol solutions. In these experiments the number of Sb-S SILAR cycles was kept two lower than the number of Sn-S sycles. Each SILAR cycle leads to a greater concentration of Sn, which is accompanied by a band gap reduction. Details of the growth procedure will be reported elsewhere. The nominal composition obtained was $Sn_xSb_yS$ with $x \approx 0.38$, $y \approx 0.28$ based on energy-dispersive X-ray spectroscopy (EDS). This is near the composition of $Sn_2Sb_2S_5$. The optical properties of the synthesized Sn-Sb-S semiconductor were characterized by transmission measurements.

Figure S8 (a) displays the transmission spectra $T(\lambda)$ of a Sn-Sb-S samples with the number of Sb-S SILAR cycles $(n) = 9$. The energy gap $E_g$ was calculated from the Tauc plot $(Ahv)^2$ vs $hv$, where $A$ is the absorbance. The intercept to the x-axis yields an approximate energy gap of $E_g = 1.35$ eV, which is consistent with our calculated result. The photovoltaic performance of the Sn-Sb-S semiconductor was investigated by measuring the power conversion efficiency (PCE) of Sn-Sb-S quantum dot-sensitized liquid junction solar cells under 0.05 sun. These cells were fabricated by assembling the Sn-Sb-S coated $TiO_2$ electrodes with a Pt counter electrode in a thin film configuration with ~ 190 $\mu$m Parafilm as a sealant and spacer. A polyiodide solution was used as the electrolyte.

The performance decreases as intensity is increased, as is usual for new solar absorber materials in this type of cell. The reason is that the synthesized nanoparticles inherently contain a large number of surface defects acting as carrier recombination centers. Under reduced light intensity, the number of photogenerated electrons decreases, leading to lower carrier recombination. Our experiments revealed a PCE of 5% under a reduced light intensity of 0.05 sun, which is a very respectable efficiency for a new solar material in initial tests. The experimental results support the theoretical predication in Figure 3 that Sn-Sb-S could be a potential solar material. It should be emphasized that further characterization of the material and optimization of it and the cells are needed before definite conclusions about the performance can be made. Details of the photovoltaic experiment will be published separately.

Compounds with the combination of band gap lower than 1.0 eV, and static dielectric constant $\varepsilon_0 > 20$ may be good candidate thermoelectric materials for waste heat recovery or other applications. We calculated the electronic fitness function, which is a measure of the extent to which a band structure has features that decouple the thermopower from the conductivity (Figure 4a,b). This is an important ingredient in achieving high thermoelectric performance.[54,62] $AgBi_3S_5$,[63] FeSbS, CoSbS,[26,64] were found to have good thermoelectric properties in previous work. In addition to these, we find that $Pb_6Bi_2S_9$, IrBiS, RhBiS, $TlBiS_2$, $Sn_5Sb_2S_9$ also may have very good thermoelectric properties at this temperature, according the EFF. It should, however, be noted that the Rh and Ir compounds would be impractical for this application due to cost. In any case, we chose the best five compounds for p- and n-type thermoelectric materials and show their Seebeck coefficients in Figure 4c,d. The band structures and PDOS are shown in Figure S2. The best materials according to the EFF are $Pb_6Bi_2S_9$ for p-type (with carrier concentration ~$5 \times 10^{18}$ $cm^{-3}$) and $TlBiS_2$ for n-type (at ~$2 \times 10^{18}$ $cm^{-3}$). As seen, both of these compounds have Seebeck coefficients 250 μV/K < $|S|$ < 400 μV/K, similar to known high performance thermoelectrics, for doping levels near the EFF maximum.

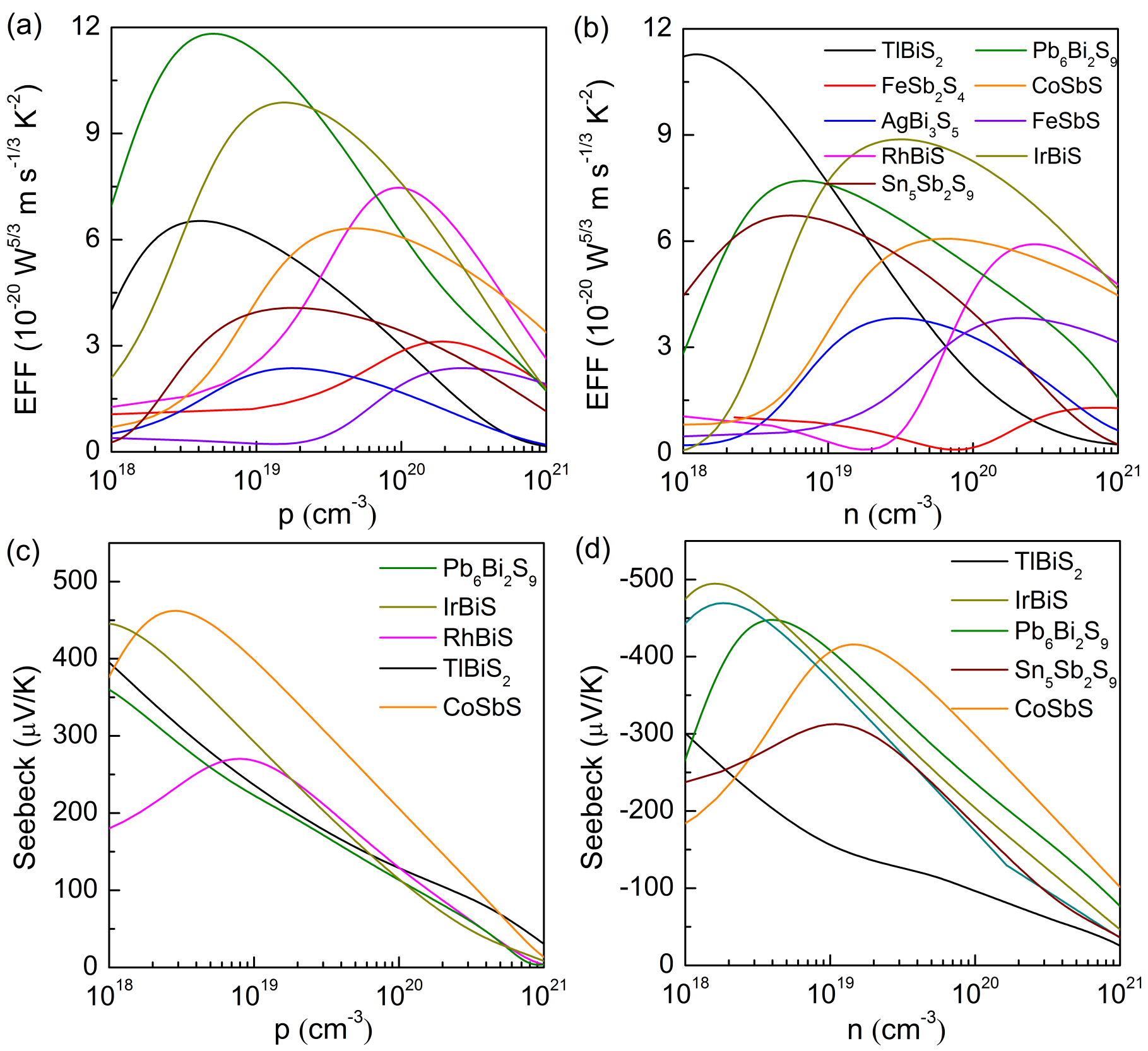

**Figure 4.** (a, b) Calculated electronic fitness function (EFF) at 800K with respect to the carrier concentration of p-type and n-type of compounds that have big electric constant and Eg < 1. (c, d) Calculated Seebeck coefficient of best five compounds from the point of view of EFF for p-type and n-type.

## Conclusions

We report a computational material search study using the dielectric properties as a screen to identify overlooked electronic materials in a family of experimentally synthesized compounds, metal–pnictogen sulfosalts (M-Sb/Bi-S). These compounds have essentially ionic electronic structures reflecting the electronegativity difference between Sb/Bi and S. Importantly, there is strong cross-gap hybridization between S p and pnictogen *p* orbitals, which leads to enhanced Born effective charges. We find that there are large dielectric constants as a consequence in structures with connected pnictogen–S sublattices. Of the 116 ternary sulfosalts that we investigated, 78 are semiconductors. The calculated band gaps of these M-Sb/Bi-S compounds cover a wide range. We find that while dielectric constant does correlate with the band gap, as might be expected, it is possible to use dielectric constant as an independent screen because there are substantial numbers of high dielectric constant materials in any band gap range, including high band gap. Finally, we identify several promising materials for optoelectronic and thermoelectric applications. We also observe a structural rule, where certain structures, i.e. those having at least one dimensionally connected pnictogen-S bonding topologies also have high dielectric constant and vice versa. We illustrate the use dielectric properties as a way of sifting through the thousands of possible semiconducting compounds, specifically the use of computed static dielectric constant based on crystal structure as a primary screen, with additional property screens depending on application. Our study suggests that rational discovery by consideration of dielectric constant is a feasible and useful approach to discover new semiconductors for electronic, optoelectronic and thermoelectric applications.

## ASSOCIATED CONTENT

### Supporting Information

Calculated absorption spectra and spectroscopic limited maximum efficiencies for several promising compounds, crystal structures and electronic band structures of the compounds with good optoelectronic or thermoelectric properties, experimental optical transmission spectrum for Sn-Sb-S nanocrystals, structure information and calculated intrinsic properties for all the ternary M-Sb/Bi-S sulfosalts studied.

### Acknowledgments

The work at Jilin University is supported by the National Natural Science Foundation of China (Grant No. 61722403 and 11674121), the National Key Research and Development Program of China (Grant No. 2016YFB0201204), and Program for Jilin University Science and Technology Innovative Research Team. Work at the University of Missouri is supported by the U.S. Department of Energy, Basic Energy Science through Award Number DE-SC0019114

# Supporting Information

for

# Dielectric Behavior as a Screen in Rational Searches for Electronic Materials: Metal Pnictide Sulfosalts

Xin He,[1] David J. Singh,[2,*] Patsorn Boon-on,[3] Ming-Way Lee[3,*] and Lijun Zhang[1,*]

[1]State Key Laboratory of Superhard Materials, Key Laboratory of Automobile Materials of MOE, and School of Materials Science, Jilin University, Changchun 130012, China

[2]Department of Physics and Astronomy, University of Missouri, Columbia, MO 65211-7010 USA

[3]Institute of Nanoscience and Department of Physics, National Chung Hsing University, Taichung 402, Taiwan

*Authors to whom correspondence should be addressed: singhdj@missouri.edu, mwl@phys.nchu.edu.tw, lijun_zhang@jlu.edu.cn

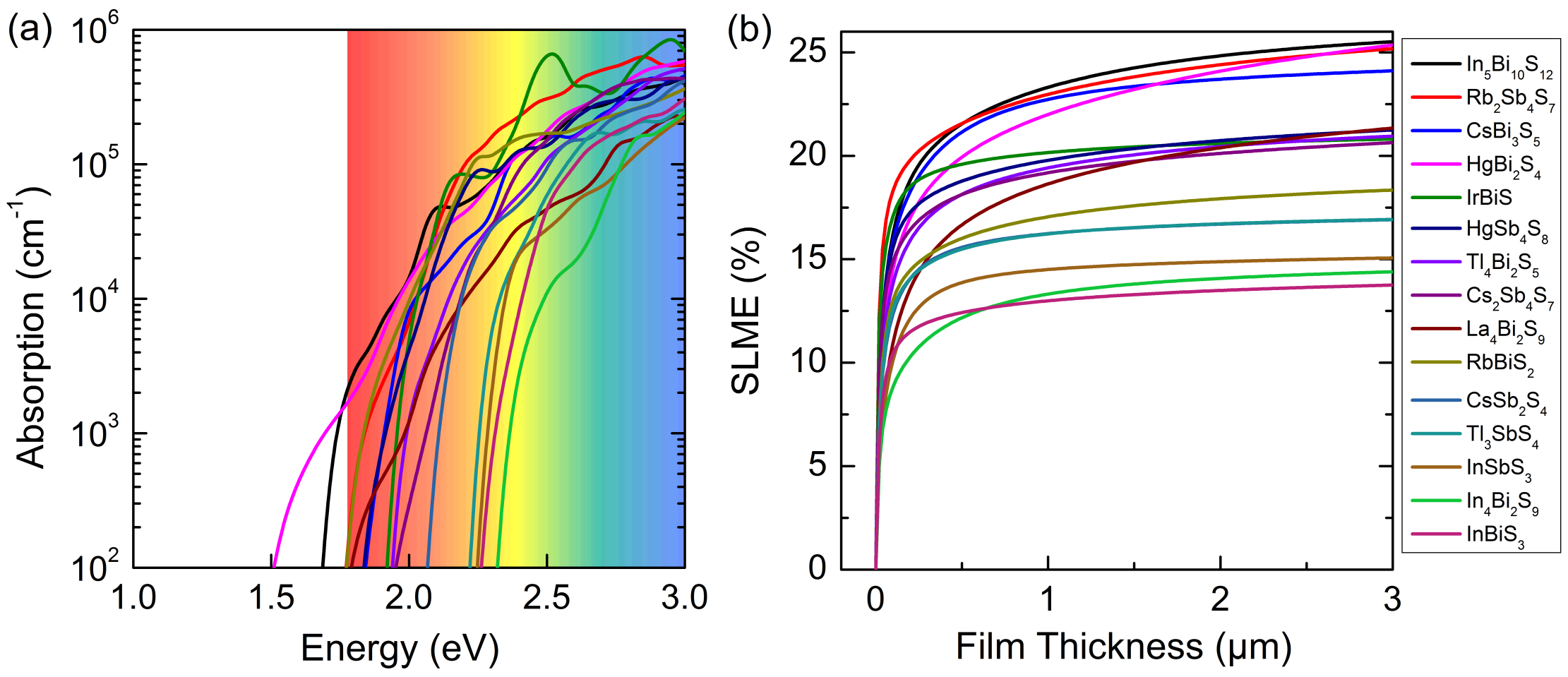

**Figure S1.** Calculated absorption spectra and "spectroscopic limited maximum efficiency (SLME) of the compounds that have electron or hole effective masses lower than the free electron mass ($m_0$), static dielectric constant $\varepsilon_0$ > 20, band gap 2.3 eV > Eg > 1.0 eV, and SLME < 25%.

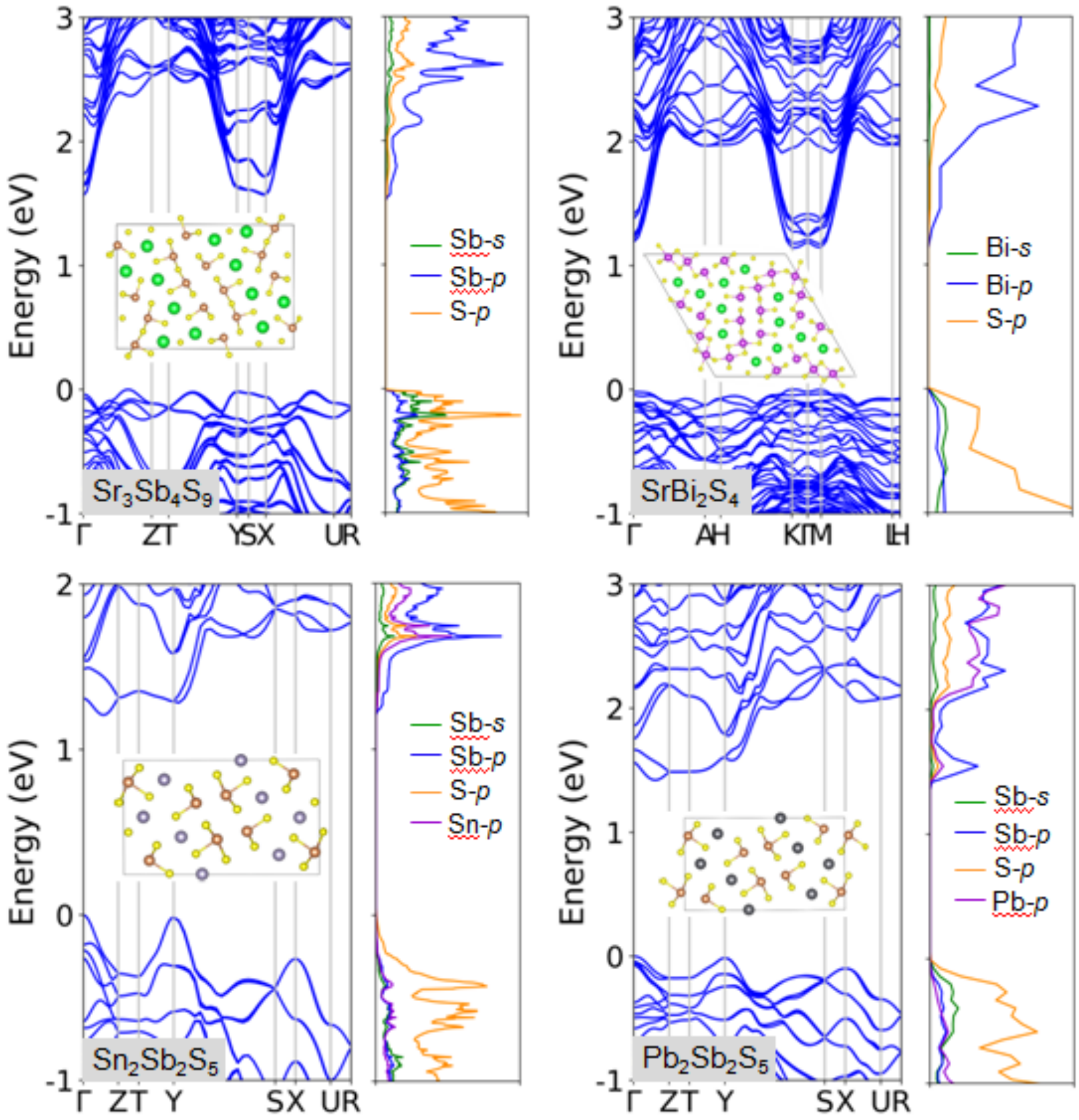

**Figure S2.** Crystal structure, band structure and projected density of states (PDOS) of the compounds having superior properties for optoelectronics (see the main text). The Bi, Sb, and S atoms in the compounds are shown in pink, brown and yellow, respectively.

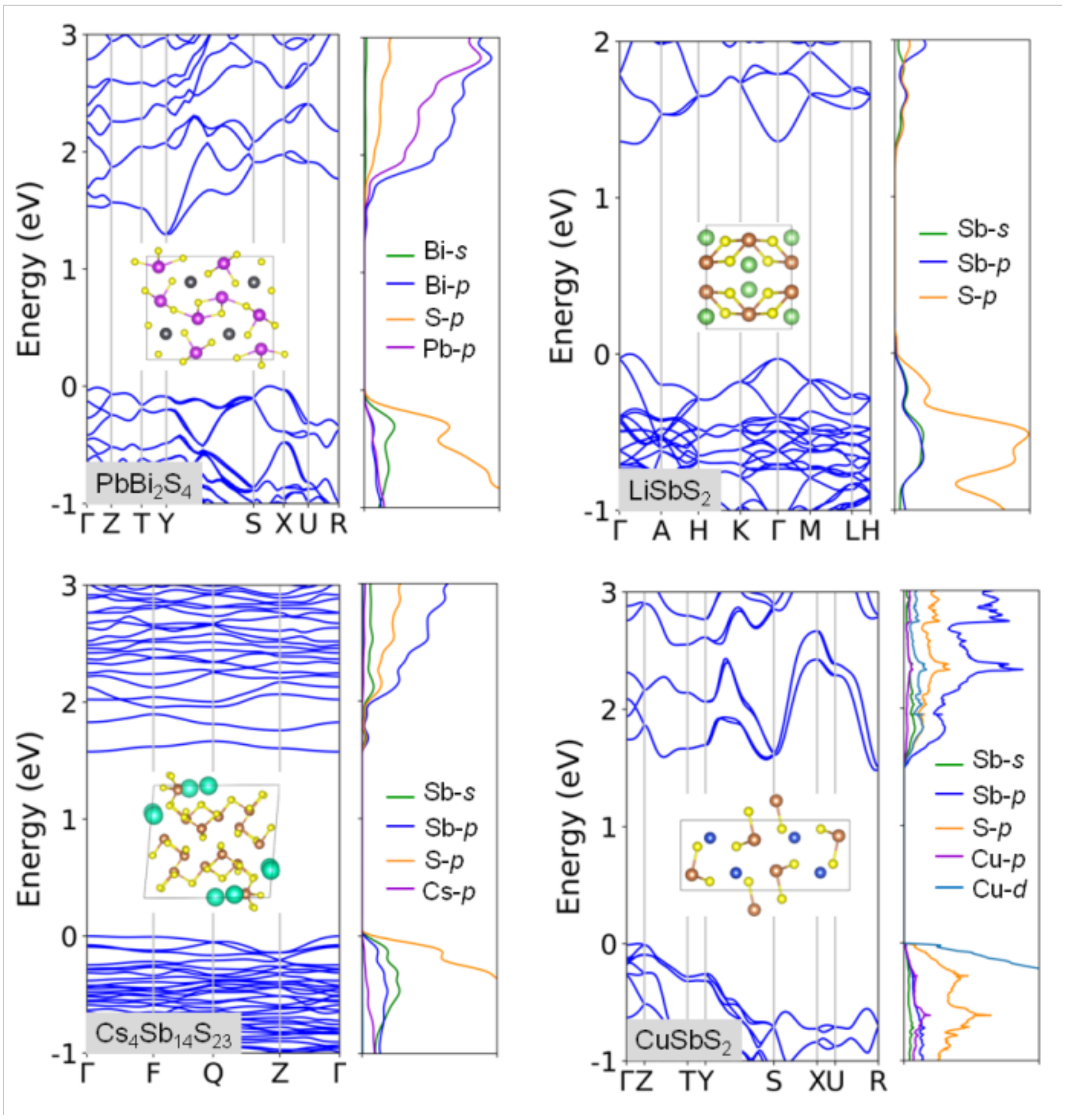

**Figure S3.** Crystal structure, band structure and projected density of states (PDOS) of the compounds having superior properties for optoelectronics (see the main text). The Bi, Sb, and S atoms in the compounds are shown in pink, brown and yellow, respectively.

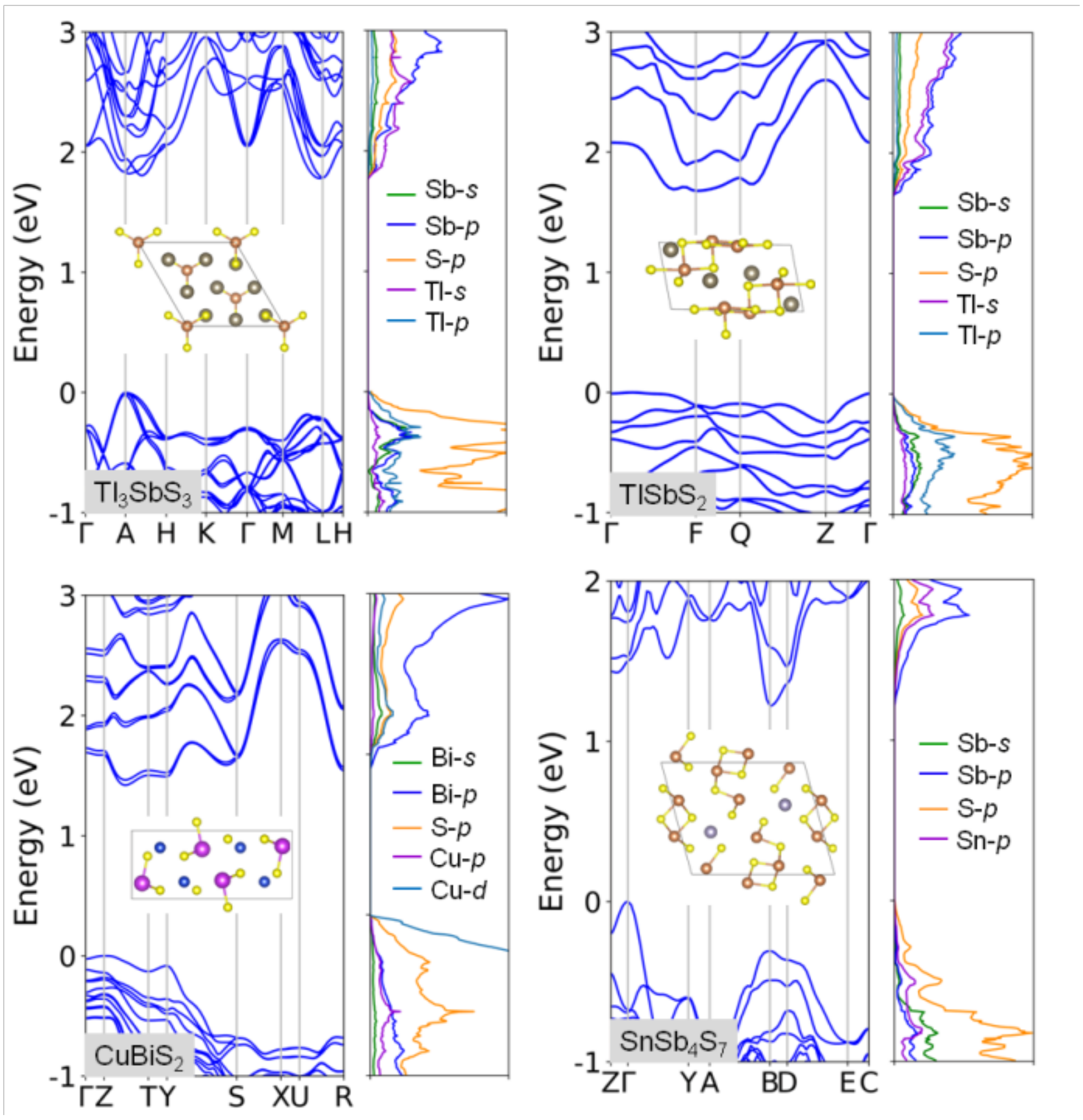

**Figure S4.** Crystal structure, band structure and projected density of states (PDOS) of the compounds having superior properties for optoelectronics (see the main text). The Bi, Sb, and S atoms in the compounds are shown in pink, brown and yellow, respectively.

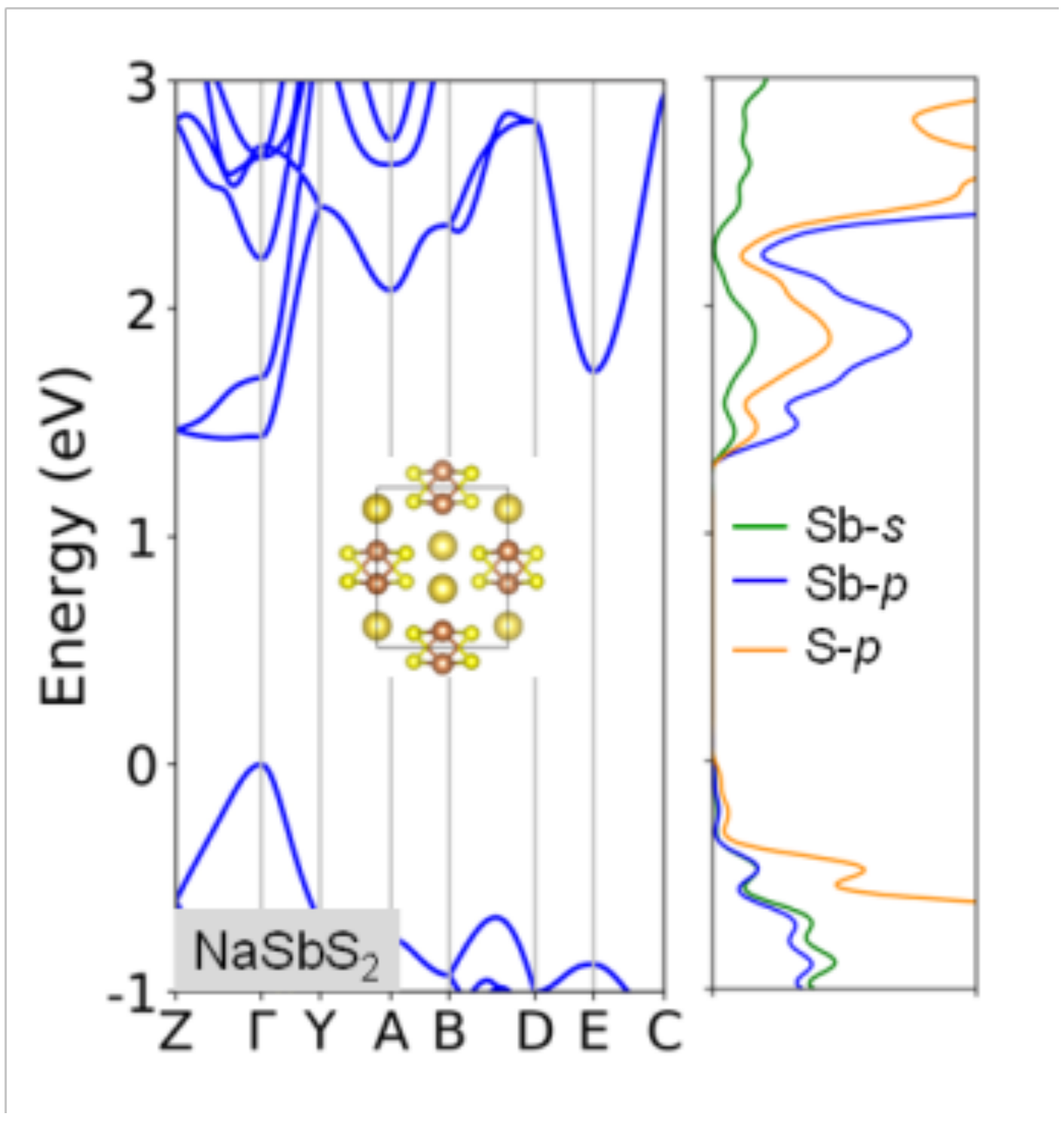

**Figure S5.** Crystal structure, band structure and projected density of states (PDOS) of the compounds having superior properties for optoelectronics (see the main text). The Sb, and S atoms in the compounds are shown in brown and yellow, respectively.

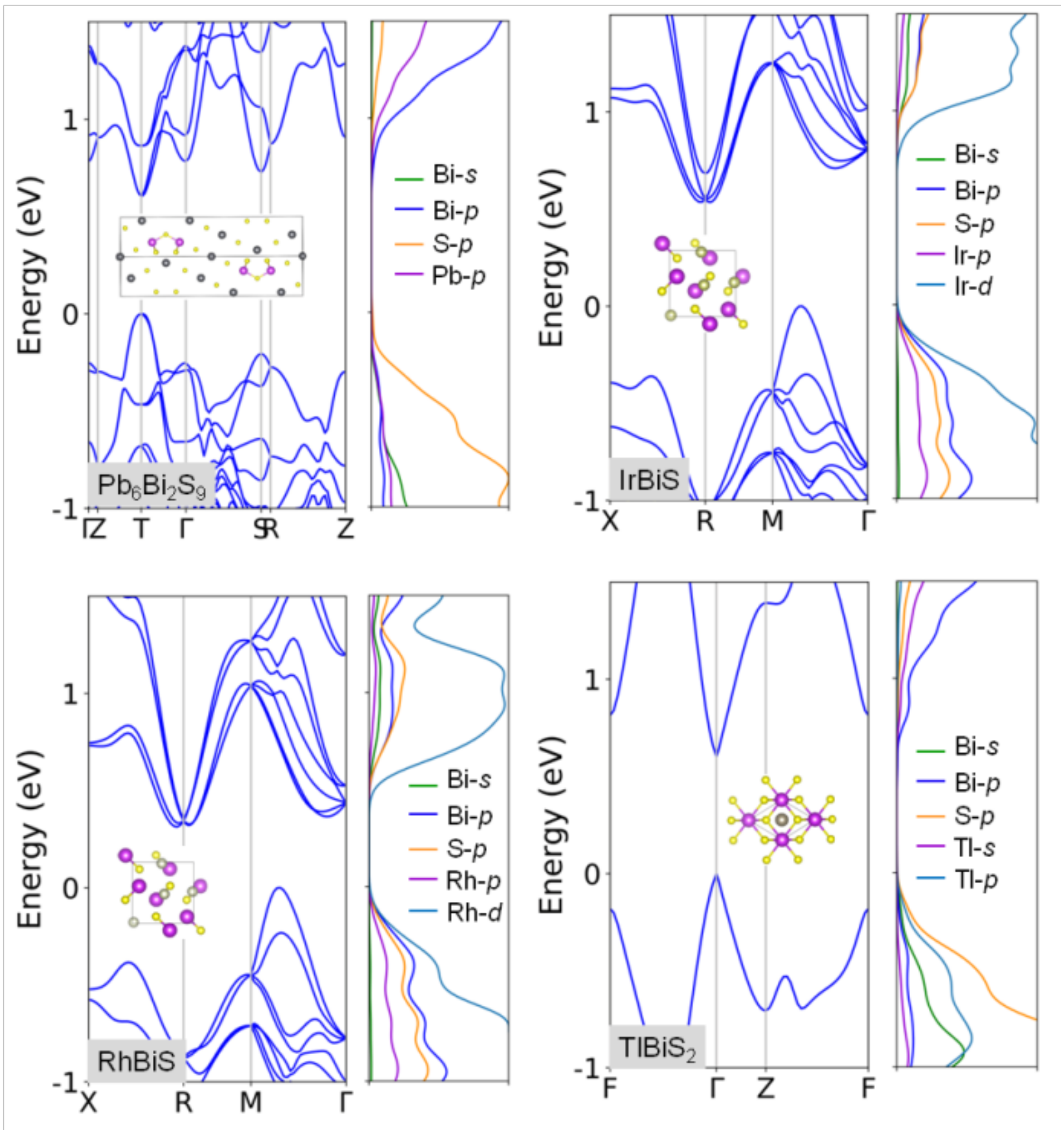

**Figure S6.** Crystal structure, band structure and projected density of states (PDOS) of the compounds having superior properties for thermoelectrics (see the main text). The Bi, Sb, and S atoms in the compounds are shown in pink, brown and yellow, respectively.

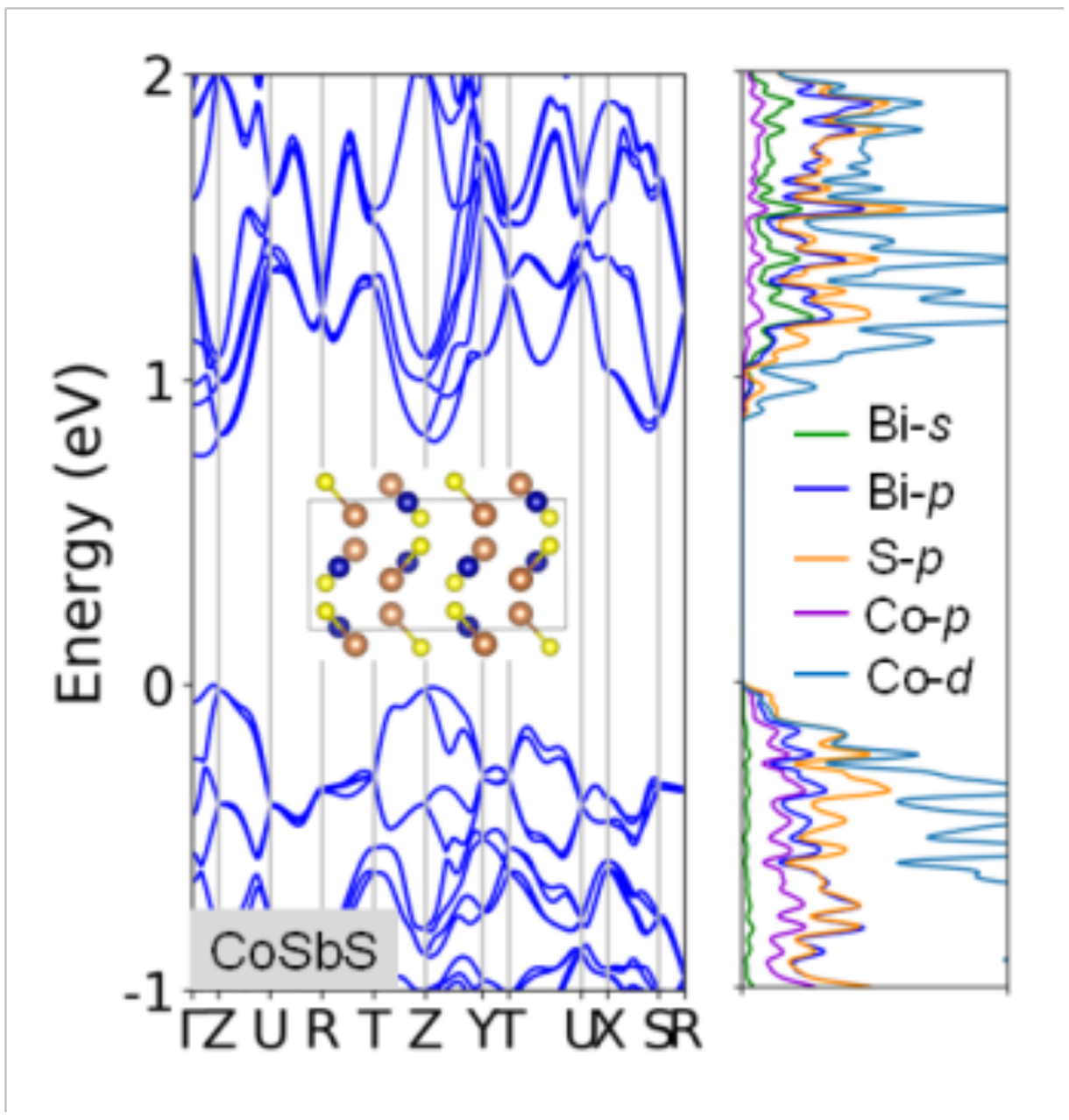

**Figure S7.** Crystal structure, band structure and projected density of states (PDOS) of the compounds having superior properties for thermoelectrics (see the main text). The Bi and S atoms in the compounds are shown in pink and yellow, respectively.

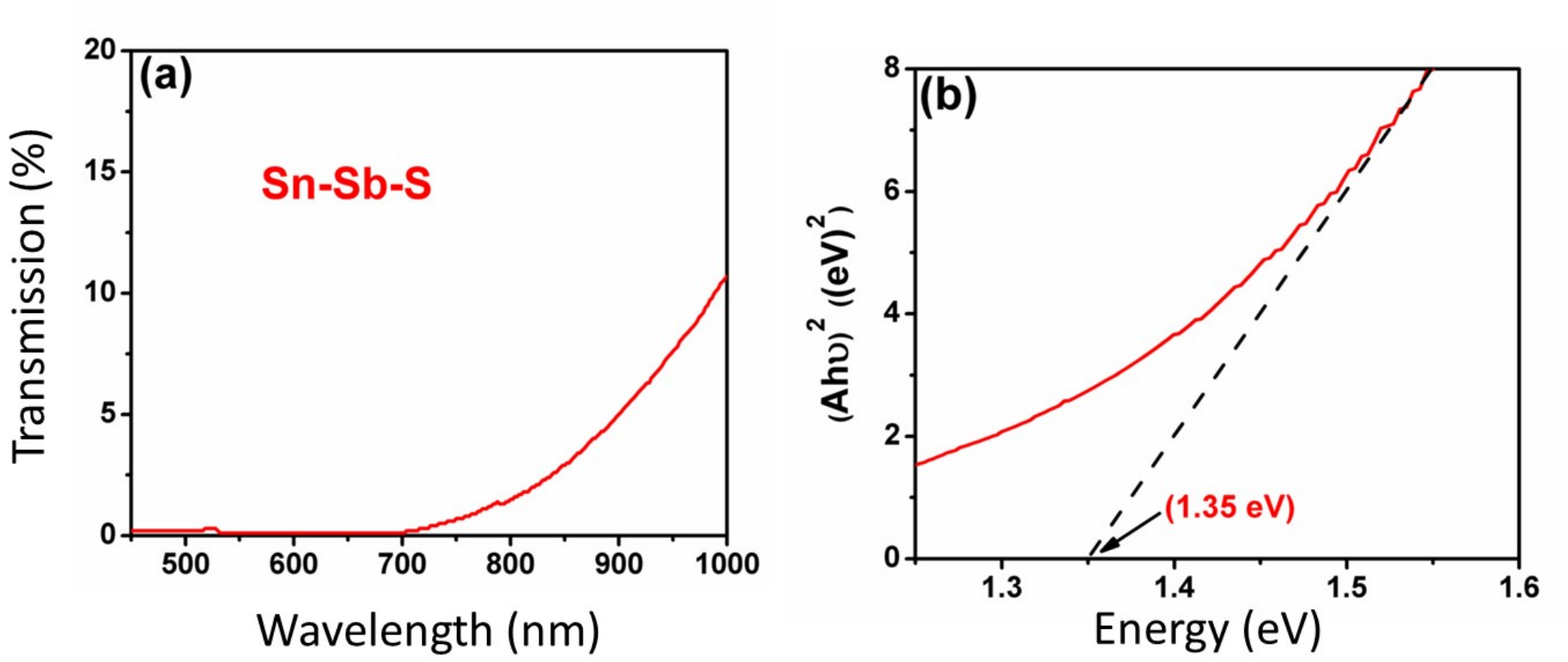

**Figure S8.** Experimental (a) transmission spectrum for the Sn-Sb-S nanocrystals ($Sn_xSb_yS$ with x~0.38, y~0.28) and (b) Tauc plot for estimation of the band gap.

**Table S1.** Lattice parameters, and calculated static dielectric constants, band gaps, Born effective charges and effective masses of electron and hole for 116 ternary sulfosalts in M-Sb/Bi-S system. Experimental gaps are shown in red.

| Compound | Structure information[1] (lattice parameters in Å) | Dielectric constant $\varepsilon_0$ | Band gap (eV) Eg (indirect) | Eg (direct) | $Z^*_{Sb/Bi}$ | $m^*_e$, $m^*_h$ ($m_0$) |
|---|---|---|---|---|---|---|
| $Te_2Bi_3S$ | space group: P-3m1 (Z=2) a=4.316, b=4.316, c=23.43 α=β=90°, γ=120° | 81.30 | 0.18 | 0.18 | 6.89 | 0.71, 1.45 |
| $Te_2Bi_2S$ | space group: R-3m (Z=3) a=4.196, b=4.196, c=29.44 α=β=90°, γ=120° | 80.23 | 0.28, 0.37[2] | 0.28 | 6.32 | 0.19, 0.16 |
| $PbBi_2S_4$ | space group: Pnma (Z=4) a=11.802, b=14.57, c=4.08 α=β=γ=90° | 79.63 | 1.29 | 1.33 | 6.14 | 0.26, 2.28 |
| $Sn_2Sb_2S_5$ | space group: Pnma (Z=4) a=19.56, b=3.938, c=11.426 α=β=γ=90° | 79.41 | 1.21 | 1.32, 1.5[3] | 5.77 | 0.54, 0.41 |
| $AgBi_3S_5$ | space group: C2/m (Z=4) a=13.35, b=4.042, c=16.439 α=β=90°, γ=90.158° | 76.37 | 0.56, 0.6[4] | 0.57 | 6.53 | 0.54, 1.11 |
| $TlBiS_2$ | space group: Pnma (Z=4) a=19.56, b=3.938, c=11.426 α=β=γ=90° | 68.46 | 0.60, 0.4[5] | 0.60 | 6.70 | 0.21, 0.21 |
| $Tl_4Bi_2S_5$ | space group: Pnma (Z=4) a=16.76, b=17.396, c=4.09 α=β=γ=90° | 60.03 | 1.86 | 1.96, 2.09[6] | 6.46 | 0.19, 0.44 |
| $SnSb_4S_7$ | space group: P21m (Z=2) a=11.331, b=3.865, c=13.94 α=γ=90°, β=105.28° | 57.16 | 1.22 | 1.42, 1.8[7] | 6.20 | 0.5, 0.45 |
| $Cs_2Sb_4S_7$ | space group: P21c (Z=4) a=11.11, b=12.27, c=11.637 α=γ=90°, β=97.6° | 49.20 | 1.89 | 1.89 | 5.02 | 0.62, 2.62 |
| $SrBi_2S_4$ | space group: P63/m (Z=12) a=b=24.925, c=4.095 α=γ=90°, β=120° | 47.16 | 1.14 | 1.14 | 6.62 | 0.48, 7.89 |
| $FeSb_2S_4$ | space group: Pnma (Z=4) a=11.42 b=14.148, c=3.76 α=γ=β=90° | 46.59 | 0.38 | 0.67 | 5.64 | 2.16, 4.53 |
| $TlSb_3S_5$ | space group: P21/c (Z=4) a=7.225, b=15.547, c=8.946 α=γ=90°, β=113.55° | 44.22 | 1.84 | 1.88 | 5.65 | 1.52, 2.41 |
| $Pb_3Sb_8S_{15}$ | space group: C2/c (Z=4) a=13.441, b=11.726, c=16.93 α=γ=90°, β=94.71° | 43.47 | 1.90 | 1.98 | 4.56 | 2.35, 2.27 |
| $RbBiS_2$ | space group: Cccm (Z=3) a=b=4.189, c=23.51 α=γ=90°, β=120° | 43.23 | 1.38, 1.36[8] | 1.77 | 6.11 | 0.27, 1.75 |
| $HgBi_2S_4$ | space group: C2/m (Z=4) a=14.17, b=4.06, c=13.99 α=γ=90°, β=118.27° | 42.51 | 1.43 | 1.43 | 6.13 | 0.28, 0.56 |
| $Pb_5Sb_8S_{17}$ | space group: C2/c (Z=4) a=13.486, b=11.86, c=19.98 α=γ=90°, β=107° | 42.05 | 1.53 | 1.53 | 6.25 | 1.35, 2.08 |
| $In_5Bi_3S_{12}$ | space group: C2/m (Z=4) a=33.13, b=3.873, c=14.413 α=γ=90°, β=91.21° | 41.84 | 1.61 | 1.70 | 6.36 | 0.92, 2.47 |
| FeSbS | space group: P21/c (Z=4) a=c=6.02, b=5.93, α=β=90°, γ=101.36° | 41.08 | 0.80 | 0.95 | 3.82 | 0.92, 1.68 |
| $CsBi_3S_5$ | space group: Pnma (Z=4) a=4.06, b=12.098, c=21.098 α=β=γ=90° | 40.79 | 1.80 | 1.81 | 6.06 | 0.51, 4.25 |
| $Pb_2Sb_2S_5$ | space group: Pnma (Z=2) a=11.355, b=19.783, c=4.042 α=β=γ=90° | 40.50 | 1.43 | 1.53 | 4.66 | 0.57, 0.89 |
| $TlSbS_2$ | space group: P-1 (Z=4) a=6.123, b=6.293, c=11.838 α=101.3°, β=98.4°, | 39.94 | 1.68 | 1.72, 1.69[5] | 5.57 | 0.69, 1.03 |
| | γ=103.21° | | | | | |
| $KSb_5S_8$ | space group: Pc (Z=4)<br>a=8.137, b=19.501, c=9.06<br>α=γ=90°, β=91.93° | 38.65 | 1.94, 1.82[9] | 1.97 | 6.43 | 1.91, 7.24 |
| RhBiS | space group: C2/m (Z=4)<br>a=11.277, b=8.37, c=7.93<br>α=γ=90°, β=133.294° | 38.21 | 0.45 | 0.81 | 4.72 | 0.87, 0.34 |
| $Sr_3Sb_4S_9$ | space group: Pna21 (Z=4)<br>a=16.579, b=24, c=4.09<br>α=β=γ=90° | 38.00 | 1.36 | 1.39 | 6.03 | 0.52, 2.68 |
| $InBiS_3$ | space group: P21 (Z=2)<br>a=9.763, b=3.786, c=6.428<br>α=γ=90°, β=88.57° | 37.55 | 2.00 | 2.26 | 6.27 | 0.43, 0.95 |
| $LiSbS_2$ | space group: R-3 (Z=18)<br>a=b=13.902, c=9.207<br>α=γ=90°, β=120° | 37.34 | 1.35, 1.36[8] | 1.35 | 1.98 | 0.8, 1.21 |
| CoSbS | space group: Pbca (Z=8)<br>a=5.842, b=5.951, c=11.666<br>α=β=γ=90° | 36.48 | 0.75 | 0.77 | 4.24 | 2.12, 1.99 |
| $Sn_5Sb_2S_9$ | space group: Pbca (Z=4)<br>a=11.437, b=11.432,<br>c=11.431 α=β=γ=90° | 35.70 | 0.95 | 0.99 | 5.81 | 0.49, 1.56 |
| $Pb_6Bi_2S_9$ | space group: Cmcm (Z=4)<br>a=13.712, b=31.21, c=4.121<br>α=β=γ=90° | 31.80 | 0.62 | 0.62 | 7.21 | 0.32, 0.36 |
| IrBiS | space group: P213 (Z=4)<br>a=b=c=6.143<br>α=β=γ=90° | 31.69 | 0.72 | 1.06 | 4.52 | 0.45, 0.37 |
| $CuBiS_2$ | space group: Pnma (Z=4)<br>a=6.143, b=3.919, c=14.528<br>α=β=γ=90° | 31.08 | 1.42 | 1.56, 1.65[10] | 6.26 | 0.48, 3.08 |
| $In_4Bi_2S_9$ | space group: P21/m (Z=2)<br>a=16.167, b=3.917, c=11.11<br>α=β=90°, γ=90.94°[11] | 30.01 | 2.27 | 2.34 | 5.63 | 0.86, 2.68 |
| $Ba_8Sb_6S_{17}$ | space group: P2/c (Z=4) a=11.41, b=13.73, c=22.53 α=β=90°, γ=90.94° | 30.26 | 1.62 | 1.62 | 4.51 | 1.21, 3.23 |
| $BaSb_2S_4$ | space group: P21/c (Z=8) a=8.985, b=8.203, c=20.602 α=β=90°, γ=101.36°[12] | 29.06 | 1.88 | 2.10 | 4.53 | 1.17, 1.17 |
| RuSbS | space group: P21/m (Z=4) a=6.18, b=6.14, c=6.19 α=γ=90°, β=111.7° | 28.4 | 1.09 | 1.23 | 3.92 | 1.23, 1.11 |
| $Sn_4Sb_6S_{13}$ | space group: C2/m (Z=4) a=24.31, b=3.915, c=23.49 α=β=90°, γ=94.05° | 28.3 | 0.68 | 0.82 | 4.22 | 0.62, 0.83 |
| $InSbS_3$ | space group: Pnma (Z=4) a=9.3, b=3.816, c=13.348 α=β=γ=90°[13] | 27.84 | 2.08 | 2.26 | 5.12 | 0.43, 1.47 |
| $CuSbS_2$ | space group: Pnma (Z=4) a=6.016, b=3.797, c=14.499 α=β=γ=90° | 27.63 | 1.48, 1.5[14] | 1.60 | 5.85 | 0.71, 2.93 |
| $La_4Bi_2S_9$ | space group: Pnma (Z=4) a=28.55, b=4.06, c=12.82 α=β=γ=90° | 27.52 | 1.68 | 1.68 | 5.42 | 0.83, 1.1 |
| $Cs_4Sb_{14}S_{23}$ | space group: P-1 (Z=2) a=11.858, b=14.164, c=14.81 α=93.19°, β=94.6, γ=110.9° | 27.49 | 1.58 | 1.58 | 4.35 | 0.85, 3.79 |
| RhSbS | space group: P213 (Z=4) a=b=c=6.027 α=γ=90°, β=111.7° | 25.9 | 1.05 | 1.3 | 3.79 | 1.54, 1.03 |
| OsSbS | space group: P21/m (Z=4) a=6.21, b=6.15, c=6.22 α=γ=90°, β=111.7° | 25.5 | 1.36 | 1.5 | 3.87 | 1.08，1.32 |
| $Tl_3SbS_4$ | space group: P-1 (Z=2) a=6.285, b=6.364, c=11.647 α=94.6°, β=98.51°, γ=103.9° | 25.00 | 2.02 | 2.24 | 3.16 | 0.41, 1.19 |
| $Tl_3SbS_3$ | space group: R3m (Z=3) a=9.519, b=9.519, c=7.364 α=γ=90°, β=120° | 23.74 | 1.78, 1.78[15] | 1.81 | 3.53 | 0.68, 0.89 |
| $NaSbS_2$ | space group: C2/c (Z=4) a=8.232, b=8.252, c=6.836 α=γ=90°, β=124.28° | 23.69 | 1.23, 1.48[8] | 1.32 | 5.70 | 0.39, 0.37 |
| $HgSb_4S_8$ | space group: C2/c (Z=8) a=30.567, b=4.01, c=21.46 α=γ=90°, β=103.39° | 22.47 | 1.57, 1.68[16] | 1.73 | 6.62 | 0.43, 0.62 |
| IrSbS | space group: P213 (Z=4) a=b=c=6.025 α=β=γ=90° | 21.45 | 1.76 | 1.94 | 4.11 | 0.84, 0.5 |
| $Rb_2Sb_4S_7$ | space group: P21 (Z=2) a=7.169, b=12.207, c=8.456 α=γ=90°, β=94.13° | 20.96 | 1.78 | 1.78 | 6.10 | 0.49, 1.36 |
| $CsSb_2S_4$ | space group: P-1 (Z=1) a=6.743, b=9.577, c=6.367 α=91.6°, β=104.8°, γ=80.57° | 20.04 | 1.78 | 2.06, 2.05[17] | 4.66 | 0.58, 1.11 |
| $AgSbS_2$ | space group: Cc (Z=8) a=12.862, b=4.41, c=13.22 α=γ=90°, β=98.63° | 19.85 | 1.61 | 1.74 | 4.93 | 0.66, 1.29 |
| $BaBi_2S_4$ | space group: P63/m (Z=12) a=b=25.272, c=4.183 α=β=90°, γ=120° | 19.63 | 1.21 | 1.28 | 4.68 | 1.44, 4.72 |
| $RbSbS_2$ | space group: P-1 (Z=4) a=b=6.473, c=6.367 α=103.6°, β=101.8°, γ=104° | 19.62 | 2.03 | 2.10 | 5.44 | 0.74, 5.67 |
| $TlSb_5S_8$ | space group: Pc (Z=4) a=8.098, b=19.415, c=9.06 α=γ=90°, β=91.96° | 19.26 | 1.72 | 1.80 | 3.98 | 1.79, 5.70 |
| $Sr_6Sb_6S_{17}$ | space group: P212121 (Z=4) a=8.287, b=15.35, c=22.87 α=γ=β=90° | 18.36 | 2.02 | 2.04 | 5.23 | 1.54, 7.56 |
| $Ag_5SbS_4$ | space group: Cmc21 (Z=4) a=7.83, b=12.45, c=8.538 α=β=γ=90° | 17.56 | 1.18 | 1.18 | 3.45 | 0.42, 1.32 |
| $CsSbS_2$ | space group: P21c (Z=4) a=7.059, b=9.784, c=7.715 α=γ=90°, β=101.44° | 16.98 | 2.43 | 2.62, 2.8[8] | 4.96 | 0.62, 2.5 |
| $Ca_2Sb_2S_5$ | space group: P21c (Z=4) a=15.07, b=5.694, c=11.378 α=γ=90°, β=110.99° | 16.76 | 1.71 | 1.71 | 4.23 | 0.3, 0.55 |
| $CsBiS_2$ | space group: P21c (Z=4) a=7.794, b=9.61, c=7.329 α=γ=90°, β=102.16°[18] | 16.55 | 2.12, 1.46[8] | 2.28 | 5.25 | 0.58, 2.42 |
| $Cu_3BiS_3$ | space group: P212121 (Z=4) a=7.723, b=10.395, c=6.72 α=β=γ=90° | 16.45 | 0.79 | 0.80, 1.20[19] 1.24[25] | 4.49 | 1.25, 0.97 |
| $KSbS_2$ | space group: C2/c (Z=4) a=8.75, b=8.98, c=6.84 α=γ=90°, β=101.6° | 15.24 | 1.68 | 1.78 | 5.52 | 0.6, 0.92 |
| $Ga_2BiS_4$ | space group: P4nnc (Z=2) a=b=7.458, c=12.032 α=β=γ=90° | 14.52 | 2.44 | 2.53 | 4.50 | 3.87, 2.41 |
| $Ba_3Sb_2S_7$ | space group: C2/c (Z=8) a=18.31, b=12.2, c=13.12 α=γ=90°, β=101.13° | 13.82 | 2.50 | 2.57 | 4.14 | 2.2, 5.19 |
| $Ag_3SbS_3$ | space group: R3c (Z=6) a=11.04, b=11.04, c=8.72 α=β=90°, γ=120° | 13.45 | 1.32 | 1.68 | 3.17 | 0.49, 0.55 |
| $Cu_3SbS_3$ | space group: P21/c (Z=8) a=7.808, b=10.23, c=13.268 α=γ=90°, β=90.31° | 13.20 | 1.13 | 1.27, 1.61[20] | 3.59 | 1.38, 0.96 |
| $Pr_8Sb_2S_{15}$ | space group:I41cd (Z=8) a=b=15.626, c=19.659 α=γ=β=90° | 12.62 | 2.62 | 2.71 | 3.02 | 1.02, 6.45 |
| $Na_3SbS_3$ | space group: P213 (Z=4) a=b=c=8.642 α=β=γ=90° | 12.51 | 3.10 | 3.17 | 2.52 | 1.24, 7.03 |
| $Nd_8Sb_2S_{15}$ | space group:I41cd (Z=8) a=b=15.58, c=19.625 α=γ=β=90° | 11.89 | 2.95 | 3.01 | 3.05 | 1.07, 5.11 |
| $Na_3SbS_4$ | space group: I-43m (Z=2) a=b=c=7.17 α=β=γ=90° | 10.92 | 2.83 | 2.83 | 3.08 | 0.86, 1.58 |
| $Li_3SbS_3$ | space group: Pna21 (Z=4) a=7.967, b=6.788, c=10.09 α=β=γ=90° | 10.91 | 2.96 | 3.03 | 3.12 | 0.61, 2.57 |
| $La_8Sb_2S_{15}$ | space group:I41cd (Z=8) a=b=15.96, c=19.855 α=γ=β=90° | 10.51 | 2.42 | 3.12 | 2.98 | 1.05, 7.43 |
| $Cs_3SbS_3$ | space group: P213 (Z=4) a=b=c=10.259 α=β=γ=90° | 10.40 | 3.49 | 3.51 | 2.18 | 15.19, 23.75 |
| $Rb_3SbS_3$ | space group: P213 (Z=4) a=b=c=9.823 α=β=γ=90° | 10.05 | 3.58 | 3.66 | 2.18 | 3.57, 18.81 |
| $K_3SbS_3$ | space group: P213 (Z=4) a=b=c=9.472 α=β=γ=90° | 9.66 | 3.52 | 3.56 | 2.21 | 3, 14.4 |
| $Cs_3SbS_4$ | space group: Pnma (Z=4) a=9.939, b=11.667, c=9.99 α=β=γ=90° | 9.45 | 3.14 | 3.14 | 2.88 | 0.79, 12.95 |
| $CsSbS_6$ | space group: P21/c (Z=4) a=5.885, b=14.41, c=11.079 α=γ=90°, β=101.69° | 9.25 | 2.22, 2.25[17] | 2.23 | 4.13 | 0.37, 7.43 |
| $Rb_3SbS_4$ | space group: Pnma (Z=4) a=9.602, b=11.161, c=9.773 α=β=γ=90° | 8.99 | 3.10 | 3.10 | 2.88 | 0.61, 4.86 |
| $Al_2BiS_4$ | space group:P4mmc (Z=2) a=b=7.492, c=11.883 α=γ=β=90° | 8.26 | 3.12 | 3.15 | 2.91 | 1.3, 5.26 |
| $K_3SbS_4$ | space group: Cmc21 (Z=4) a=10.712, b=11.289, c=7.705 α=β=γ=90° | 7.80 | 3.19 | 3.19 | 2.89 | 0.87, 2.45 |
| $RbBi_3S_5$ | space group: Pnnm (Z=4) a=4.161, b=12.902, c=18.478 α=γ=β=90° | -- | -- | -- | -- | -- |
| $Mo_2SbS_2$ | space group: P21/m (Z=2) a=6.51, b=3.18, c=9.355 α=γ=90°, β=105.44° | -- | -- | -- | -- | -- |
| MoSb2S | space group: C2/m (Z=169) a=36.16, b=6.38, c=6.34 α=γ=90°, β=95.09° | -- | -- | -- | -- | -- |
| $Mo_6BiS_8$ | space group: R-3 (Z=3) a=b=9.194, c=11.325 α=γ=90°, β=92.15° | -- | -- | -- | -- | -- |
| $MnSb_2S_4$ | space group: Pnma (Z=4) a=11.459, b=3.823, c=14.351 α=β=γ=90° | -- | -- | -- | -- | -- |
| $Mo_6SbS_8$ | space group: R-3 (Z=3) a=b=9.122, c=11.282 α=γ=90°, β=120° | -- | -- | -- | -- | -- |
| $Pb_3Bi_2S_6$ | space group: C2/m (Z=4) a=13.51, b=4.085, c=20.649 α=γ=90°, β=92.15° | -- | -- | -- | -- | -- |
| $Pb_4Sb_6S_{13}$ | space group: P1 (Z=2) a=16.56, b=17.69, c=3.98 α=91.1°, β=96.46, γ=96.77°[21] | -- | -- | -- | -- | -- |
| $Pb_4Sb_4S_{11}$ | space group: Pbam (Z=2) a=15.01, b=15.56, c=4.068 α=γ=β=90°[22] | -- | -- | -- | -- | -- |
| $Pb_7Sb_8S_{19}$ | space group: C2/c (Z=4) a=13.628, b=11.94, c=21.285 α=β=90°, γ=90.92°[23] | -- | -- | -- | -- | -- |
| $AgBi_6S_9$ | space group: C2/m (Z=2) a=13.83, b=4.04, c=14.72 α=β=90°, γ=97.5° | -- | -- | -- | -- | -- |
| $Zr_5Sb_3S$ | space group: P63/mcm (Z=2) a=b=8.4265, c=5.899 α=β=90°, γ=120° | -- | -- | -- | -- | -- |
| $CrSbS_3$ | space group: Pnma (Z=4)<br>a=8.665, b=3.619, c=12.872<br>α=γ=β=90° | -- | -- | -- | -- | -- |
| $Rh_3Bi_2S_2$ | space group: C2/m (Z=4)<br>a=11.278, b=8.37, c=7.93<br>α=γ=90°, β=133.29°[24] | -- | -- | -- | -- | -- |
| $Ni_3Bi_2S_2$ | space group: Pmma (Z=1)<br>a=5.545, b=5.731, c=4.052<br>α=β=γ=90° | -- | -- | -- | -- | -- |
| NiSbS | space group: P213 (Z=1)<br>a=b=c=5.934<br>α=β=γ=90° | -- | -- | -- | -- | -- |
| PdSbS | space group: P213 (Z=4)<br>a=b=c=6.185<br>α=γ=β=90° | -- | -- | -- | -- | -- |
| $Pd_3Bi_2S_2$ | space group: I213 (Z=4)<br>a=b=c=8.3097<br>α=β=γ=90° | -- | -- | -- | -- | -- |
| PtSbS | space group: P213 (Z=1)<br>a=b=c=6.174<br>α=β=γ=90° | -- | -- | -- | -- | -- |
| $Cu_4Bi_5S_{10}$ | space group: C2/m (Z=2)<br>a=17.539, b=3.931,<br>c=12.847<br>α=γ=90°, β=108° | -- | -- | -- | -- | -- |
| $Cu_4Bi_7S_{12}$ | space group: C2/m (Z=2)<br>a=17.52, b=3.912, c=15.24<br>α=γ=90°, β=101.26° | -- | -- | -- | -- | -- |
| $Cu_4Bi_4S_9$ | space group: Pnma (Z=4)<br>a=11.589, b=32.05, c=3.951<br>α=γ=β=90° | -- | -- | -- | -- | -- |
| $Cu_{12}Sb_4S_{13}$ | space group: I-42m (Z=4)<br>a=b=c=10.391<br>α=γ=β=90° | -- | -- | -- | -- | -- |
| $Cu_3SbS_4$ | space group: I-42m (Z=2)<br>a=b=5.385, c=10.754<br>α=γ=β=90° | -- | -- | -- | -- | -- |
| $Ce_4Bi_2S_9$ | space group: Pnma (Z=4) a=28.52, b=4.06, c=12.8 α=γ=β=90° | -- | -- | -- | -- | -- |
| $Nd_4Bi_2S_9$ | space group: Pnma (Z=4) a=28.52, b=4.02, c=12.72 α=γ=β=90° | -- | -- | -- | -- | -- |
| $Eu_3Bi_4S_9$ | space group: Pnma (Z=4) a=17.34, b=4.08, c=24.7 α=γ=β=90° | -- | -- | -- | -- | -- |
| $EuBi_2S_4$ | space group: Pnma (Z=4) a=12.5, b=4.07, c=14.03 α=γ=β=90° | -- | -- | -- | -- | -- |
| $Eu_2BiS_4$ | space group: Pnma (Z=4) a=11.579, b=14.523, c=4.089 α=γ=β=90° | -- | -- | -- | -- | -- |
| $Eu_3Sb_4S_9$ | space group: Pnma (Z=4) a=16.495, b=23.843, c=4.03 α=γ=β=90° | -- | -- | -- | -- | -- |
| $EuSb_2S_4$ | space group: Pnma (Z=4) a=11.22, b=3.98, c=13.86 α=γ=β=90° | -- | -- | -- | -- | -- |
| $Eu_6Sb_6S_{17}$ | space group: P212121 (Z=4) a=8.236, b=15.237, c=22.724 α=γ=β=90° | -- | -- | -- | -- | -- |
| GdSbS | space group: P21c (Z=4) a=4.27, b=17.3, c=6.12 α=γ=90°, β=139.7° | -- | -- | -- | -- | -- |
| $Dy_4Bi_2S_9$ | space group: Pnma (Z=4) a=28.50, b=4.04, c=12.76 α=γ=β=90° | -- | -- | -- | -- | -- |
| $YdBi_2S_4$ | space group: Pnma (Z=4) a=12.47, b=4.05, c=14.00 α=γ=β=90° | -- | -- | -- | -- | -- |
| $YdSb_2S_4$ | space group: Pnma (Z=4) a=11.2, b=3.94, c=13.82 α=γ=β=90° | -- | -- | -- | -- | -- |
| $Yb_3Sb_4S_9$ | space group: Pnma (Z=4)<br>a=16.46, b=3.98, c=23.66<br>α=γ=β=90° | -- | -- | -- | -- | -- |
| YbSbS | space group: P21c (Z=4)<br>a=4.25, b=17.25, c=6.09<br>α=γ=90°, β=139.9° | -- | -- | -- | -- | -- |